\documentstyle[aps,preprint,floats,epsf]{revtex}
\begin{document}
\newcommand {\be}{\begin{equation}}
\newcommand {\ee}{\end{equation}}
\newcommand {\bea}{\begin{eqnarray}}
\newcommand {\eea}{\end{eqnarray}}
\newcommand {\nn}{\nonumber}

\draft

%
%
%

\title{Lightly Doped {\it t-J} Three-Leg Ladders \\
- an Analog for the Underdoped Cuprates}

\author{ 
T.M. Rice$^{(1)}$, Stephan Haas$^{(1)}$, Manfred Sigrist$^{(1)}$ \cite{sigr}, and
Fu-Chun Zhang$^{(2)}$}
\address{$^{(1)}$ Theoretische Physik,
ETH-H\"onggerberg, CH-8093 Z\"urich, Switzerland\\
$^{(2)}$ Department of
Physics, University of Cincinnati, Cincinnati, OH 45221 }  
\date{\today}
\maketitle

\begin{abstract}

The three-leg ladder has one odd-parity and two even-parity channels.
At low doping these behave quite differently. Numerical calculations
for a $t-J$ model show that the initial phase upon hole doping has two
components - a conducting Luttinger liquid in the odd-parity channel,
coexisting with an insulating (i.e. undoped) spin liquid phase in the
even-parity channels. This phase has a partially truncated Fermi surface
and violates the Luttinger theorem. This coexistence of conducting
fermionic and insulating paired bosonic degrees of freedom is similar
to the recent proposal of Geshkenbein, Ioffe, and Larkin
(Phys. Rev. B{\bf 55}, 3173 (1997)) for the underdoped spin-gap normal
phase of the cuprates. 
A mean field approximation is derived which has many similarities to the
numerical results. One difference however is an induced hole pairing in
the odd-parity channel at arbitrary small dopings, similar to that proposed
by Geshkenbein, Ioffe, and Larkin for the two-dimensional case. 
At higher dopings, we propose that a quantum phase transition will occur
as holes enter the even-parity channels, resulting in a Luther-Emery
liquid with hole pairing with essentially d-wave character. In the mean
field approximation a crossover occurs which we interpret as a reflection
of this quantum phase transition deduced from the numerical results.

\end{abstract}

\vskip2pc

\section{Introduction}

The properties of electrons confined to ladders with various numbers of
legs have been investigated by many groups in the past few years
\cite{dagotto}. The
reason for this interest lies both in the unusual properties of the ladder
systems and the possibility to realize ladder structures in the cuprates,
but also in the insight these systems give to the full two-dimensional
problem of a square lattice. Two different approaches have been taken. One is
based on the weak-coupling limit and uses renormalization group methods to
analyze the multi-leg ladders
\cite{schulz,khveshchenko,arrigoni,balents,kimura,lin}. 
A very complete analysis of this type for
N-leg ladders in the Hubbard model has recently been given by Lin,
Balents and Fisher\cite{lin}. 

A second approach is more numerical and examines the strong-coupling limit
described mostly by the $t$-$J$ model. Recent progress on the loop algorithm for
Monte Carlo calculations has allowed large undoped systems described 
by a Heisenberg model to be investigated down to very low temperatures. However,
when doped holes are introduced, the fermion sign problem prevents one 
from using
this method, and other methods must be employed, e.g. using Lanczos techniques
to diagonalize relatively
small systems\cite{troyer,hayward} 
or the new density matrix renormalization group
method (DMRG) to obtain the groundstate of large systems\cite{white}.

In this paper we examine the case of the lightly doped $t-J$ three-leg
ladder. This case is specially
interesting because in a certain sense it combines the
contrasting properties of a single chain\cite{ogata,bares}
and a two-leg ladder\cite{troyer}. 
These behave quite differently, both when undoped or when lightly doped,
so it is of great interest to follow the evolution of the three-leg ladder
in this regime. In particular, the evolution of the Fermi surface as
the Mott insulating state is approached is very different in the different
transverse channels. As we shall discuss further below, this leads to a 
region where the Fermi surface is truncated in two channels, but remains
in one channel - a behavior which clearly violates the Luttinger theorem.
We shall argue that this presence of transverse channels which
behave quite differently helps us to make inferences for the limit of the
two-dimensional plane which of course can be represented as the limit of very 
many channels or patches on the Fermi surface. The lightly doped limit of the
three-leg ladder can serve as a simpler analog for the underdoped
spin gap region of the cuprates, which has attracted so much attention, 
if one assumes that the patches or channels near $(\pm \pi, 0)$ and
$(0,\pm \pi )$ are truncated and
show a spin gap, while those near $(\pm \pi/2, \pm \pi/2)$
are gapless\cite{marshall}.

The $t$-$J$ three-leg ladder Hamiltonian is given by
\bea
H = &-&t \sum_{j,\sigma } \sum_{\nu=1}^3 P (c^{\dagger }_{j,\nu,\sigma }
c_{j+1,\nu,\sigma } + H.c.)P \nn \\
&-&t' \sum_{j,\sigma } \sum_{\nu=1}^2 P (c^{\dagger }_{j,\nu,\sigma }
c_{j,\nu+1,\sigma } + H.c.)P \nn \\
&+& J \sum_{j} \sum_{\nu=1}^3 ( {\bf S}_{j,\nu} \cdot {\bf S}_{j+1,\nu} 
- \frac{1}{4}n_{j,\nu}\cdot n_{j+1,\nu} ) \nn \\
&+& J' \sum_{j} \sum_{\nu=1}^2 ( {\bf S}_{j,\nu} \cdot {\bf S}_{j,\nu+1} 
- \frac{1}{4}n_{j,\nu}\cdot n_{j,\nu+1} ),
\eea
where $j$ runs over $L$ rungs, $\sigma(=\uparrow,\downarrow)$, and $\nu$ are
spin and leg indices. The $t$-$J$ three-leg
ladder is sketched in Fig. 1. The first two
terms are the kinetic energy ($P$ is a projection operator which prohibits
double occupancy), and the last two exchange couplings $J$ ($J'$) act along
the legs (rungs). Periodic or antiperiodic boundary conditions (PBC, APBC)
are used along the legs.

\begin{figure}[htb]
\centerline{\epsfxsize 12cm
\epsffile{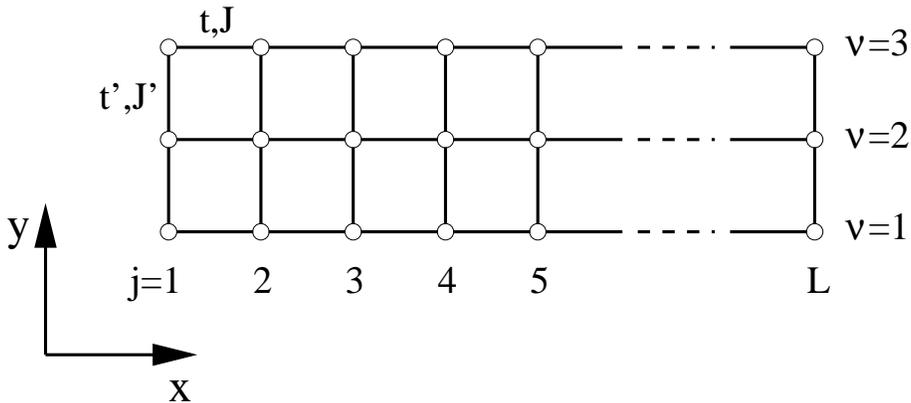}
}
\vspace{5mm}
\caption{The $t$-$J$ ladder with three legs and L rungs. The couplings along the
legs are $t$ and $J$, and those along the rungs $t'$ and $J'$.}
\end{figure}

The paper is organized as follows. In the next section we recapitulate
briefly the known results in the undoped limit described by a Heisenberg
Hamiltonian. Then in section 3 we discuss the case of a single doped hole,
using the results of a Lanczos diagonalization and also the earlier DMRG
results by White and Scalapino\cite{white}. 
The low-energy properties are described
by a single Luttinger liquid channel in contact with an insulating spin liquid (ISL).
Next in section 4 we consider states with two and more holes. 
In the presence of a finite hole density, there are two possibilities -
either all the holes repel each other and enter the single Luttinger liquid 
channel, or at some density, the other channels are also populated with holes.
Since doping a resonating valence bond (RVB) spin liquid leads to a 
Luther-Emery liquid\cite{troyer}, 
this will cause a qualitative change in the physical
properties. 
There will be a
critical hole density, $\rm \delta_c$, which controls the transition 
between the low-density 
phase with only a Luttinger liquid, in contact with a spin liquid, to the case 
with both Luttinger and Luther-Emery liquids. In section 5 we develop a mean
field approach to the $t$-$J$ three-leg ladder. In this mean field 
approximation, as we shall see, the critical density, $\rm \delta_c$, 
is not finite but
arbitrarily small, whereas the numerical results give a finite $\rm \delta_c$
for values of $J/t \sim 0.5$. 
In the final section we draw some conclusions and discuss the
relationship to the planar two-dimensional case.

\section{Undoped Case: the Heisenberg Three-Leg Ladder}

In this section we recapitulate briefly the known results for three-leg ladders
in the Heisenberg model. In this case the fermion sign problem does not occur,
and very accurate quantum Monte Carlo calculations have been carried out by
Frischmuth {\it et al.}\cite{frischmuth}, 
and by Greven {\it et al.}\cite{greven}. The low-energy properties
can be mapped onto an effective single S=1/2 antiferromagnetic (AF) 
Heisenberg chain model, although the starting model has three spins per
set of ladder rungs. This arises because the additional spin degrees
of freedom have a spin gap, and enter with an excitation energy
$\sim J'$.
The spinon velocity of the effective model
in fact is hardly changed from the value of the 
nearest neighbor Heisenberg chain with exchange constant $J$. However the 
energy scale parameter which controls the spinon-spinon interactions and the
size of logarithmic corrections to the uniform spin susceptibility at finite
temperatures is greatly changed. Frischmuth {\it et al.}\cite{frischmuth}
interpret
this result in terms of a Heisenberg model for the effective chain
with longer range and unfrustrated
effective interactions which act to enhance the AF correlations.

The key point however is the fact that the low-energy Hilbert subspace is 
greatly reduced so that not three chains but only a single chain model is 
active at low energies, 
or in other words only a single
S=1/2 degree of freedom per rung of three spins remains in the low-energy 
region. The remaining spin degrees of 
freedom are gapped in the same way as the spin liquid of a two-leg
ladder.

\begin{figure}[htb]
\vspace{5mm}
\centerline{\epsfxsize 10cm
\epsffile{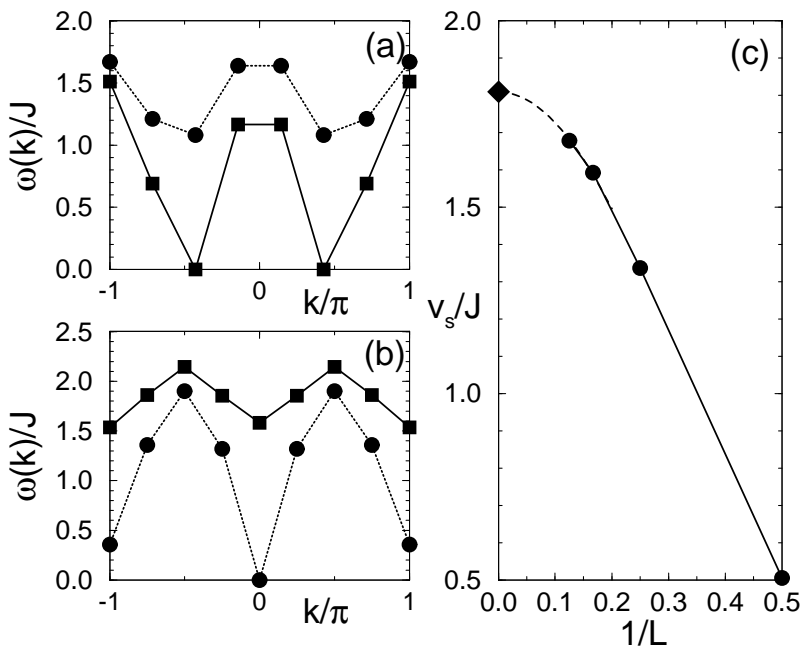}
}
\vspace{5mm}
\caption{(a) and (b) Spinon 
excitation spectra for the isotropic ($3\times 7$)- and
($3\times 8$)-site Heisenberg ladders.
The circles correspond to the even parity and the squares
to the odd parity channels respectively. 
(a) Single spinon excitation spectrum
(S = 1/2) in a ($3\times 7$)-cluster. 
(b) Two-spinon excitation spectrum
(S = 1) in a ($3\times 8$)-cluster.
(c) Spinon-velocity, $\rm v_s$, in  (3$\times$L)-Heisenberg ladders. 
$\rm v_s$ extrapolates with high accuracy to the 
value 1.81J (diamond), obtained by QMC calculations [13].}
\end{figure}

This behavior is clearly observed in the numerical Lanczos diagonalization
of ($3\times 7$)- and ($3\times 8$)-clusters, shown in Fig. 2.
As Bares {\it et al.}\cite{bares}
emphasized in their study of the
exactly solvable supersymmetric $t$-$J$ model in one dimension, the spinon and 
holon dispersions can be directly obtained in the limit of vanishing hole 
density by examining the groundstate of an appropriately chosen finite chain.
Thus if one takes the case of one undoped chain with an odd number of sites and
PBC, then the groundstate for each total wavevector along the chain gives 
the dispersion of a single spinon. This groundstate manifold has a total 
spin quantum number S=1/2. For
the present case of a three-leg ladder we show the
dispersion for a ($3\times 7$)-sample with PBC in Fig. 2(a).
The single spinon with S=1/2 has odd parity with respect to reflection
about the center leg and disperses with a bandwidth $\sim \frac{\pi}{2}J$ and
minima at $\pm \frac{\pi}{2}$. The corresponding one-spinon 
excitation spectrum of a 7-site single Heisenberg chain is shown in Fig. 3(a).
To estimate finite size effects, the calculation was repeated for a 19-site
Heisenberg chain (Fig. 3(c)).
We observe that the qualitative features, i.e. the positions of extrema and 
the overall bandwidth, are already present in the smaller cluster. This is
important since in the exact diagonalization study of ($3\times$L)-ladders,
we are restricted to relatively small (L$\leq$8) systems.

\begin{figure}[htb]
\vspace{5mm}
\centerline{\epsfxsize 10cm
\epsffile{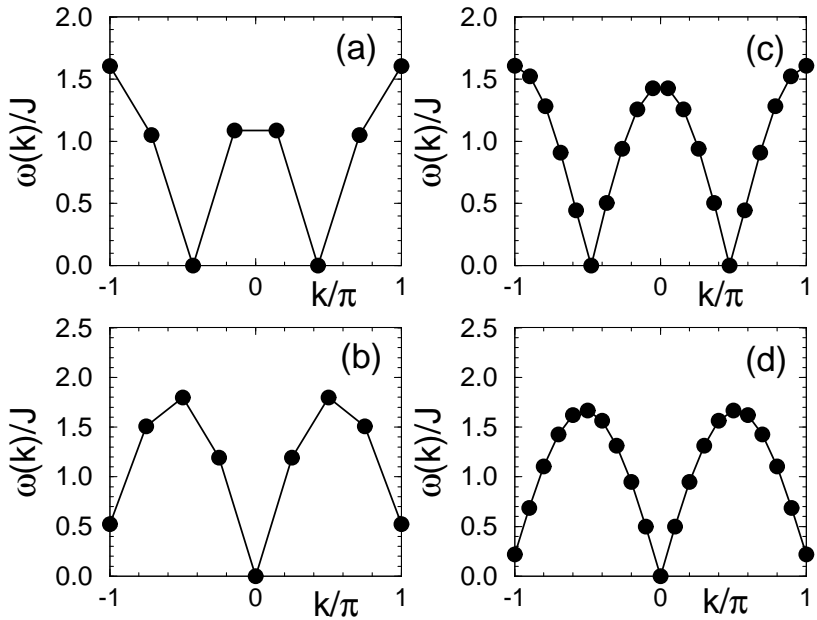}
}
\vspace{5mm}
\caption{Spinon excitation spectra for 7- and 8-site Heisenberg chains.
(a) Single spinon excitation spectrum
(S = 1/2) in a 7-site chain.
(b) Two-spinon excitation spectrum
(S = 1) in an 8-site chain. 
(c) Same as (a), but for a 19-site chain.
(d) Same as (b), but for a 20-site chain.}
\end{figure}

For the ($3\times 7$)-ladder single spinon dispersion shown in Fig. 2(a),
there appears an additional even-parity band, corresponding to the gapped
spin degrees of freedom. It is separated from the odd-parity band by a 
gap $\sim J$, and its bandwidth is much smaller than that of the low-energy 
band. 
For the case of an even number of rungs (as shown in Fig. 2(b) for the case
of ($3\times 8$)), the low-lying spin excitations now have even parity with 
respect to reflection about the center leg, and involve pairs of spinons
with S=1 and groundstate wavevectors k=$\pm \frac{\pi}{2} \pm \frac{\pi}{2}
= 0, \pm \pi$. The finite size gap at k=$\pm \pi$ vanishes in the thermodynamic
limit. Again, a comparison of the low-energy two-spinon band with that of
the corresponding 8-site single Heisenberg chain (Fig. 3(c)) shows 
qualitative (almost quantitative) agreement, supporting our conclusion that 
the low-energy degrees of freedom of the ($3\times$L)-ladders can be mapped
onto effective single chains.

Finally, we analyze the finite size scaling behavior of the spinon
velocity, $\rm v_s = \frac{\partial \omega}{\partial k}|_{k=0}$, for the
three-leg Heisenberg ladders (Fig. 2(c)).
In the finite systems, we approximate the
derivative by $\rm v_s \approx \frac{\omega(2\pi /L) - \omega(0)}
{2\pi /L}$ (L even). Using L=2, 4, 6, and 8, we obtain with the approximate
scaling form, $\rm v_s(L) = v_s(\infty ) + a L^{-2} + b L^{-4}$, 
a bulk value of $\rm v_s (\infty ) \simeq 1.81J$, which is in excellent
agreement ($\sim 1 \% $) with that obtained by recent QMC calculations on 
clusters with up to 600 spins\cite{frischmuth0}.

\section{Single Hole in a Three-Leg Ladder}

We start the study of effects of doping by recalling the single electron
non-interacting
bandstructure. This takes the form of three overlapping bands (in the isotropic
case $t = t'$). These can be classified according to their parity
under the reflection operation ($R$) about the center leg. The two
even-parity bands we denote as bonding ($b$) and anti-bonding ($ab$), and they
have the dispersion relation
\bea
\epsilon_{b,ab} (k) = \mp \sqrt{2} t' - 2 t \cos{k},
\eea
while the odd-parity band (or non-bonding band) has the form
\bea
\epsilon_{nb}  (k) = - 2 t \cos{k}.
\eea
At half-filling, the chemical potential is $\mu = 0$, and all three bands are 
partially filled. The Fermi surface consists of three pairs of Fermi points
($\pm {\rm k}_{{\rm F},\lambda}, \lambda = ab, nb, b$), arising from each band
(or channel).

\begin{figure}[htb]
\centerline{\epsfxsize 10cm
\epsffile{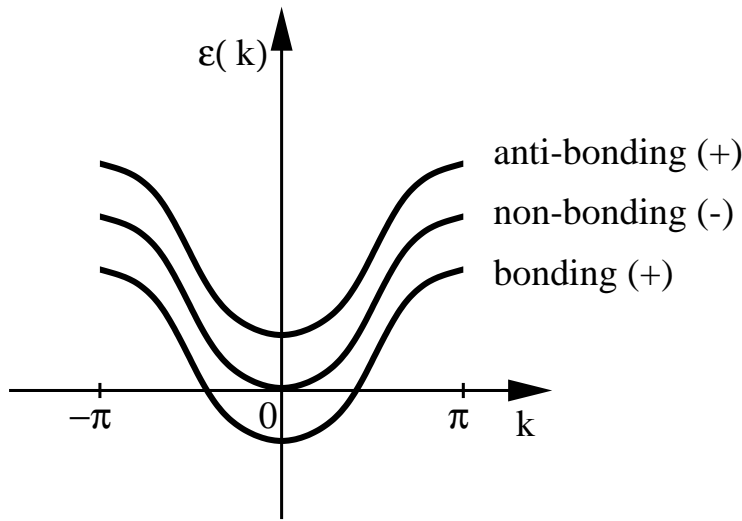}
}
\vspace{2mm}
\caption{Schematic plot of the non-interacting one-electron bandstructure of a
($3\times L$)-ladder. The parities of the bands with respect to reflection 
about the center leg are indicated by $(\pm)$. The splitting between the bands
is $\sqrt{2}t'$.}
\end{figure}

We consider the case of a single hole, and we start with
a ($3\times 7$)-sample. The 
groundstate is now a total singlet (S=0) and non-degenerate.
The wavevector dependence 
determines the single
holon dispersion, again in a single chain effective model.
In Fig. 5(a), the results are shown for a
parameter value $J/t=0.5$. The holon energy has absolute
minima at k=$\pm\pi$ and a local
minimum at k=0. This latter 
behavior is consistent with a finite size effect. For
example in Fig. 5(c) we show the holon dispersion calculated
in a 7-site single chain $t-J$ model
where a similar behavior is obtained. This in turn reflects the doubling
of the period in the single Heisenberg chain dispersion, when the chain
length goes to infinity.

\begin{figure}[htb]
\vspace{5mm}
\centerline{\epsfxsize 10cm
\epsffile{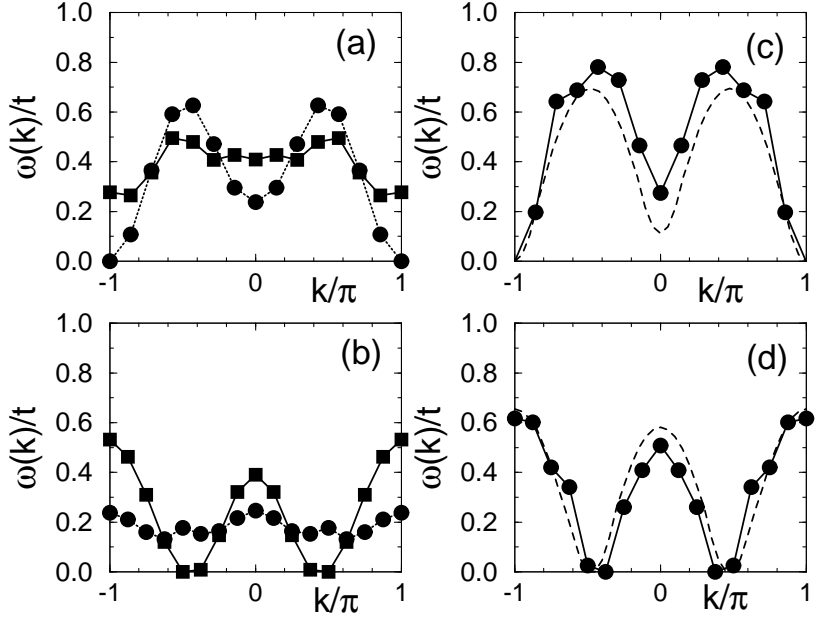}
}
\caption{Hole 
excitation spectra for the isotropic ($3\times 7$)- and
($3\times 8$)-site $t-J$ ladders, and for 7- and 8-site $t-J$ chains.
In (a) and (b), the circles correspond to the even parity and the squares
to the odd-parity channels respectively.
(a) Single holon
($\rm  \delta  = 1/21$, S = 0) in a ($3\times 7$)-cluster.
(b) Single hole excitation spectrum
($\rm  \delta = 1/24$, S = 1/2) in a ($3\times 8$)-cluster.
(c) Filled circles: 
single holon excitation spectrum
($\rm \delta  = 1/21$, S = 0) in a 7-site chain.
Dashed line: same in a 19-site chain.
(d) Filled circles:
single hole excitation spectrum
($\rm \delta = 1/24$, S = 1/2) in an 8-site chain.
Dashed line: same in a 20-site chain.}
\vspace{5mm}
\end{figure}

The three-leg ladder has reflection symmetry with respect to the center leg.
The eigenvalues of the corresponding operator, $R$, are $\pm 1$, i.e. even
(odd) parity under reflection. The groundstate for the undoped ladder with
(3$\times$L) legs (L odd) has parity -1.
The groundstate manifold of the same ladders with one hole
has parity +1.
Therefore we associate parities +1 and -1 with a single holon and
spinon respectively. It follows that the parity of a single hole which is the 
product of those parities is -1, i.e. the hole goes into the band with odd 
parity with respect to $R$. 
This is the middle non-bonding band with odd parity. The
interpretation is clear: the two bands with even parity 
in the $t-J$ model combine to form a 
spin liquid which is insulating (ISL), 
and the initial doped holes go into the
odd-parity band (or channel), and form a single Luttinger liquid (LL). So
the non-interacting
bandstructure Fermi surface is truncated from three sets of Fermi points 
to a 
single Luttinger liquid in the odd channel. Note that this Fermi surface
truncation implies a form of spin pairing and a reduction of the low-energy
spin degrees of freedom, but it does not imply the formation of Cooper pairs.
Note also that this Fermi surface truncation is not due to a breaking of
translational symmetry along the ladder since the spin-spin correlations in
the ISL are purely short ranged.

\begin{figure}[htb]
\centerline{\epsfxsize 10cm
\epsffile{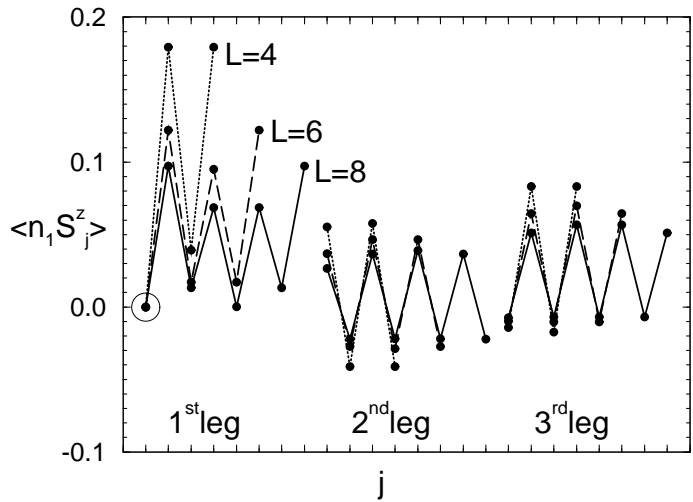}
}
\caption{Spin-charge correlations for one hole in (3$\times$L)
isotropic $t-J$ clusters with
L = 4, 6, 8, and $J/t = 0.5$. The holon is fixed at the site indicated by
the circle.}
\end{figure}

White and Scalapino have performed
DMRG calculations on large samples ($3\times 16$)
with open boundary conditions\cite{white}. 
They find upon doping a single hole that the
spin and charge densities are well separated, indicating the decay of a single 
hole into a separated spinon and holon as expected for a Luttinger liquid. 
The Lanczos results for smaller clusters show similar results. In Fig. 6,
the results for the spin-charge correlation are shown. If we add one hole
to a (3$\times$L) cluster with L even, then the groundstate has total spin
S=1/2, and odd parity. Therefore if we take the state with $S^z = +1/2$,
we can calculate the local value of $S^z$ at each site
for configurations with
the holon fixed at the origin. In Fig. 6, we show these results for L=4, 6,
and 8.
The results show clearly that the spin is distributed over the whole cluster,
and agree nicely with those of White and Scalapino\cite{white}. 
This behavior is 
fully consistent
with a single Luttinger liquid interpretation.

\begin{table}
\caption{Energies (in units of $t$) of the ($3\times 6$)-site 
isotropic $t-J$ ladder with antiperiodic boundary conditions and
$J/t=0.5$.}
\vspace{2mm}
\begin{tabular}{||c|c|c||c|c|c||c|c|c||} 
\multicolumn{3}{||c||}{0 holes } &
\multicolumn{3}{c||}{1 hole } & 
\multicolumn{3}{c||}{2 holes } \\ \hline \hline
k & $R$ & energy & k & $R$ & energy & k & $R$ & energy \\ \hline \hline 
0 & +1  & -9.017603 & $\pi/6$ & +1 & -10.24582 
& 0 & +1  & -11.80680 \\ \hline
$\pi/3$ & +1 & -8.393170 & $\pi/2$ & +1 & -10.25131
& $\pi/3$ & +1 & -11.42219 \\ \hline
$2\pi/3$ & +1 & -8.409188 & $5\pi/6$ & +1 & -10.25673
& $2\pi/3$ & +1 & -11.49704 \\ \hline
$\pi$ & +1 & -9.243230 & & &
& $\pi$ & +1 & -11.69586 \\ \hline
0 & -1  & -8.450508 & $\pi/6$ & -1 & -9.987594
& 0 & -1  & -11.59084 \\ \hline
$\pi/3$ & -1  & -8.211651 & $\pi/2$ & -1 & -10.40757
& $\pi/3$ & -1  & -11.57428 \\ \hline
$2\pi/3$ & -1 & -8.224553 & $5\pi/6$ & -1 & -10.18534
& $2\pi/3$ & -1 & -11.53402 \\ \hline
$\pi$ & -1 & -8.403831 & & &
& $\pi$ & -1 & -11.54535 \\ 
\end{tabular}
\end{table}

The Lanczos method allows one to examine also the low-lying excited states.
This in turn raises the question of the energy gap between the Luttinger
liquid with odd parity
and the next channel with even parity. This latter channel should
correspond to 
placing the hole in the spin liquid. From the experience with the two-leg ladder
one expects that doping the spin liquid will lead to a Luther-Emery liquid 
with a bound spin-charge distribution for a single hole.
Therefore if our expectation is correct, we should expect the lowest-lying
excited states with even parity to show a very different spin-charge 
distribution. In Fig. 7, we show the corresponding distributions for the 
lowest eigenstates with even and odd parities of clusters with L=6. The
eigenvalues are given in Table 1. The 
difference is clear. In the odd parity channel the energy dispersion is large,
the minimum energy lies at k=$\pm \pi/2$, corresponding to a half-filled 
Luttinger liquid. 
In the even parity channel, 
the energy dispersion is much less, and the minimum lies at values
k=$\pm 5\pi/6$ which we interpret as a hole entering the anti-bonding
even parity band. As can be seen from Fig. 7, the instantaneous spin-charge
distribution for this minimum energy even parity state is quite different,
and shows a clear binding of the spin and charge. This behavior is fully
consistent with that of a single hole in a Luther-Emery liquid and with the
interpretation of a spin liquid in the even parity channels. Also shown in
Fig. 7 are the instantaneous spin-charge distributions for the set of lowest
excited states with fixed wavevector and parity. These show intermediate
behavior which we interpret as arising from (attractive) interactions
between a spinon and a holon at wavevectors away from the Fermi surface.

\begin{figure}[htb]
\vspace{5mm}
\centerline{\epsfxsize 11cm
\epsffile{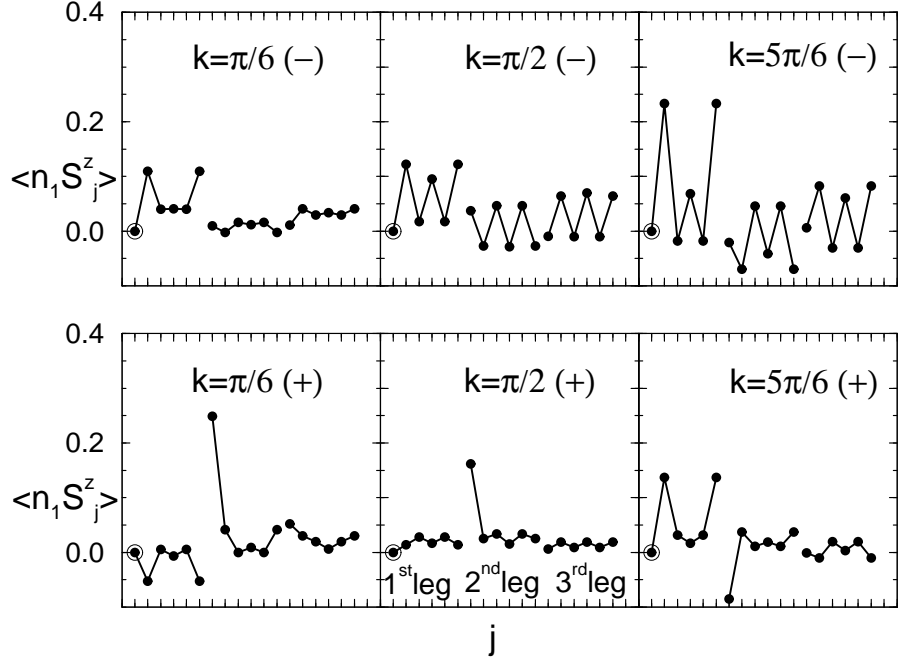}
}
\vspace{2mm}
\caption{Spin-charge correlations for one hole in a (3$\times$6)-site
isotropic $t-J$ cluster with
$J/t = 0.5$. The holon is fixed at the site indicated by
the circle.}
\end{figure}

An important quantity is the single hole
energy gap between the odd and even parity
channels. The value we obtain in the (3$\times$6)-cluster  
is $\approx 0.15t$ (or $0.30J$) - a value which is smaller than that of the 
spin gap in the ISL ($\approx J$).

To summarize, the results for a single hole show that the minimum energy 
is in the non-bonding channel, and this channel forms a single Luttinger
liquid. The even channels are gapped, forming an insulating spin liquid (ISL),
and the minimum energy for a single hole in these channels lies higher in
energy.

\section{Two and More Holes}

We begin with the case of two holes. White and Scalapino have performed DMRG
calculations for (3$\times$L)-samples with a value of $J/t=0.5$\cite{white}. 
They
find that the holes do not bind and in fact repel each other, so that the most 
likely configuration has the two holons as widely separated as possible,
consistent with the open boundary conditions used in their study.
This behavior can be immediately interpreted as that of two holes doped in
a single Luttinger liquid (LL), while the remaining transverse channels form
an ISL, consistent with the discussion given above. In view of the finite
energy gap between the even and the odd parity states, there will be a finite 
density range within which this LL + ISL phase remains stable. 
We then conclude that the hole 
density in the White-Scalapino calculations ($\rm \delta = 1/24$) lies within 
this density range. 

The next issue is to determine the critical hole density, $\rm \delta_c$,
that limits the stability of this LL + ISL phase. At first sight it would 
seem straightforward to use the single hole energy gap between odd and 
even parities to determine the critical value for the hole chemical 
potential, $\rm \mu_c$, and thus $\rm \delta_c$: $\rm \mu_c = \mu (\delta_c)$.
However, we expect the even parity channels to evolve into a Luther-Emery
liquid (LE) when doped, in analogy to the doped two-leg ladder. As a 
result, $\rm \mu_c$ will be determined by the energy of the two-hole 
bound state of the LE rather than the single-hole energy gap. 

In Fig. 8, we show the instantaneous hole-hole correlation function 
for various two-hole states of a (3$\times$6)-cluster with APBC 
($J/t$=0.5). These boundary conditions
were chosen since they give a non-degenerate
(i.e. closed shell) groundstate for non-interacting electrons\cite{footnote1}.
The groundstate has even parity and a total wavevector of k=0.
The instantaneous
hole-hole correlation function (Fig. 8(a))
shows the maximum weight at the largest
rung-rung separation possible in this small cluster, but with both holes 
preferentially on the same edge. This behavior is similar to that found
by White and Scalapino, 
and leads us to interpret this as a groundstate 
with LL + ISL character. Note that the hole density here ($\rm \delta = 1/9$),
is larger than that considered by White and Scalapino. 

\begin{figure}[htb]
\vspace{2mm}
\centerline{\epsfxsize 11cm
\epsffile{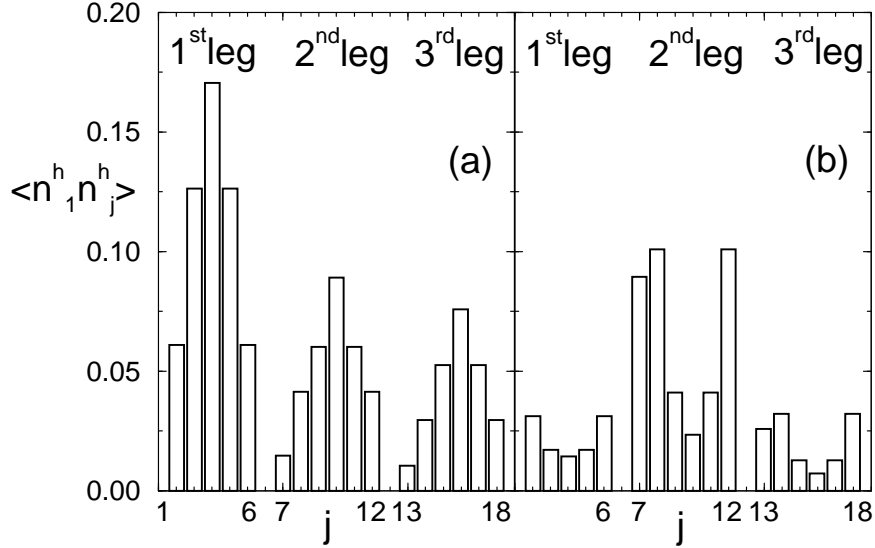}
}
\vspace{2mm}
\caption{Instantaneous hole-hole correlations for two holes in a 
(3$\times$6)-site
isotropic $t-J$ cluster with
$J/t = 0.5$. The first hole is fixed at the origin (j=1), not
shown in the figure. (a) groundstate, k=0. (b) excited state, k=$\pi$.
The sites on the first (outer) leg are labeled $\rm 1 \leq j \leq 6$,
on the second (center) leg $\rm 7 \leq j \leq 12$, and on the third
(outer) leg $\rm 13 \leq j \leq 18$.}
\end{figure}

\begin{table}
\caption{Average hole densities on the central and the outer legs in
the ($3\times 6$)-site
isotropic $t-J$ ladder with two holes and
$J/t=0.5$.}
\vspace{2mm}
\begin{tabular}{||c|c|c|c||}
k & $R$ & $\rm <n^h>_{out}$ & $\rm <n^h>_{central}$ \\ \hline \hline
0 & +1  & 1.050053  
& 0.899894  \\ \hline
$\pi/3$ & +1 & 0.791384  
& 1.417232 \\ \hline
$2\pi/3$ & +1 & 0.809848  
& 1.380304 \\ \hline
$\pi$ & +1 & 0.794342  
& 1.411316 \\ \hline
0 & -1  & 1.170123  
& 0.659754 \\ \hline
$\pi/3$ & -1  & 0.995565  
& 1.00887 \\ \hline
$2\pi/3$ & -1 & 0.978532  
& 1.042936 \\ \hline
$\pi$ & -1 & 0.884392  
& 1.231216 \\
\end{tabular}
\end{table}

There should however be an excited state of the cluster, also with even parity
and S=0, which has the two holes in the LE. This state should have a total
wavevector of $\rm k = \pi$, i.e. the same value as the undoped system
\cite{footnote2}.
In fact, such a state appears in the cluster
with an excitation energy of 0.111$t$ above the groundstate. The instantaneous
hole-hole correlations in this excited state (Fig. 8(b)) 
are also quite different
from the groundstate, with a maximum for the next-nearest-neighbor separation,
indicating that the two holes are bound in this state. Therefore we identify 
this state as that with two holes in a LE formed from the even parity 
transverse channels (i.e. bonding and anti-bonding bands). This identification
is confirmed if we look at the average hole density on the central vs. the
outer legs. The results quoted in Table 2 show a marked increase in the hole
density on the central leg in the LE state. 

The binding energy of the two holes in the LE state can be estimated from the
energy difference to add the two holes {\it together} in the LE state 
($\rm E^{LE}_2$) and {\it separately} in the LE state ($\rm E^{LE}_1$).
This difference is an estimate of the binding energy,
$\rm E_b = 2 E^{LE}_1 - E^{LE}_2$ = -2.028$t$ + 2.453$t$ = 0.425$t$.
Note however that this value may well be a considerable overestimate
because of finite size effects (see below).

We now turn to the estimate of the chemical potential, $\rm \mu (\delta)$. In
view of the small size of the clusters (i.e. mostly 3$\times$6 sites) we
need to take care about finite size corrections. In particular, in small 
clusters there is generally a large energy gain for total singlet states, 
and if one calculates $\mu$ through single hole additions one ends up
subtracting the energies of states with total spin S=0 and S=1/2. 
In order to avoid this, we use two-hole additions since then the total spin 
quantum number remains unchanged; i.e. $\rm \mu (\delta) = 
\frac{1}{2} [ E_G (N_h +1) - E_G (N_h -1) ]$. The result is shown in Fig. 9.
The curve of $\rm \mu (\delta)$ rises initially with $\rm \delta$, consistent
with a repulsive interaction between holes in the LL + ISL phase. Our
previous estimate of the excitation energy of the LE state with $\rm N_h=2$
allows us to determine the chemical potential rise needed for holes to
enter the LE phase at 0.055$t$. From Fig. 9 we estimate then for the
critical hole density a value of $\rm \delta_c \simeq 0.13$. Note that this 
value is calculated using a relatively small cluster size, and it is not
possible to estimate the finite size corrections. Therefore this value
of $\rm \delta_c$ is an estimate whose accuracy is hard to predict. 

\begin{figure}[htb]
\vspace{2mm}
\centerline{\epsfxsize 11cm
\epsffile{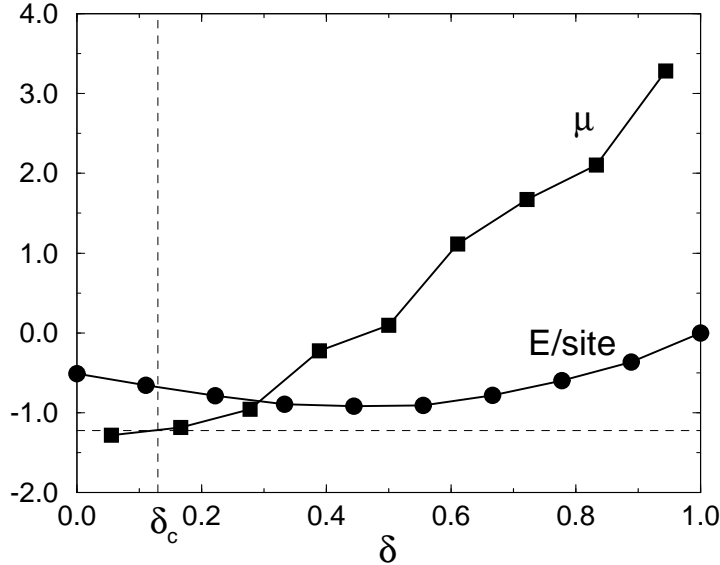}
}
\vspace{2mm}
\caption{Energy per site (circles) and chemical potential (squares)
for a (3$\times$6)-site
isotropic $t-J$ ladder with 
$J/t$=0.5. The critical hole density, $\rm \delta_c$,
beyond which the holes enter the LE
phase is indicated by a dashed line.}
\end{figure}

We conclude that the LL + ISL phase remains stable over a finite hole density
range, 0$<\rm \delta<\rm \delta_c$, and in this range there is no hole pairing 
although a large fraction of the spin degrees of freedom are gapped in the
ISL. Beyond this density range, 
$\rm \delta>\rm \delta_c$, there is a LE channel in
contact with the LL channel. The presence of holes in a LE channel
leads to hole pairing and 
dominant superconducting or charge density wave correlations as in the case
of the lightly doped two-leg ladder.

In order to further explore the low-density LL + ISL phase, we calculated
the single particle spectral function, $\rm A_{\lambda}(k,\omega)
= - \frac{1}{\pi} Im[G_{\lambda}(k,\omega - i\eta)]$, in the groundstate
of the (3$\times$6)-cluster with two holes. Here, $\lambda$ labels the
linear combinations of c-operators, corresponding to the non-interacting
shell structure shown in Fig. 5:
$c_{j,b,\sigma} = \frac{1}{2}[c_{j,1,\sigma} + \sqrt{2}c_{j,2,\sigma} +
c_{j,3,\sigma}]$, 
$c_{j,nb,\sigma} = \frac{1}{\sqrt{2}}[c_{j,1,\sigma} - 
c_{j,3,\sigma}]$, 
$c_{j,ab,\sigma} = \frac{1}{2}[c_{j,1,\sigma} - \sqrt{2}c_{j,2,\sigma} +
c_{j,3,\sigma}]$. 
The results are displayed in Fig. 10 for (3$\times$6) ladders with APBC
and PBC.
The energy region below the chemical potential corresponds to a transition
to three-hole states. In a small cluster, such as the (3$\times$6) under study 
here, the finite size effects are large. Thus to interpret the spectrum, 
one should keep in mind the non-interacting shell structure. We begin
with the case of APBC. The two-hole
LL + ISL state has then 16 electrons which in a non-interacting state would
occupy the states $b$($\pm \frac{\pi}{6}, \pm \frac{\pi}{2}$), 
$ab$($\pm \frac{\pi}{6}$), and $nb$($\pm \frac{\pi}{6}$).
Through the strong correlation interaction the state with electron pairs
in the $b$($\pm \frac{5\pi}{6}$) state will be admixed. In this way we can
interpret the electron removal (or photoemission) part of the spectrum with 
the strong peaks to inserting holes into the $b$($\pm \frac{\pi}{2},\pm
\frac{5\pi}{6}$) and $ab$($\pm \frac{\pi}{6}$) even parity channels.
The state with an added hole at $b$($\pm \frac{\pi}{6}$) is far removed from
the Fermi level, and so it is strongly broadened. Turning to the $nb$-channel,
we see that the main weight is at ($\pm \frac{\pi}{6}$) as expected, but
this is again broadened due to many-body effects. The resulting 
photoemission spectrum then shows signs of a Fermi surface in all three
channels, and with $\rm k_F$-values in line with our expectations.

\begin{figure}[htb]
\vspace{5mm}
\centerline{\epsfxsize 11cm
\epsffile{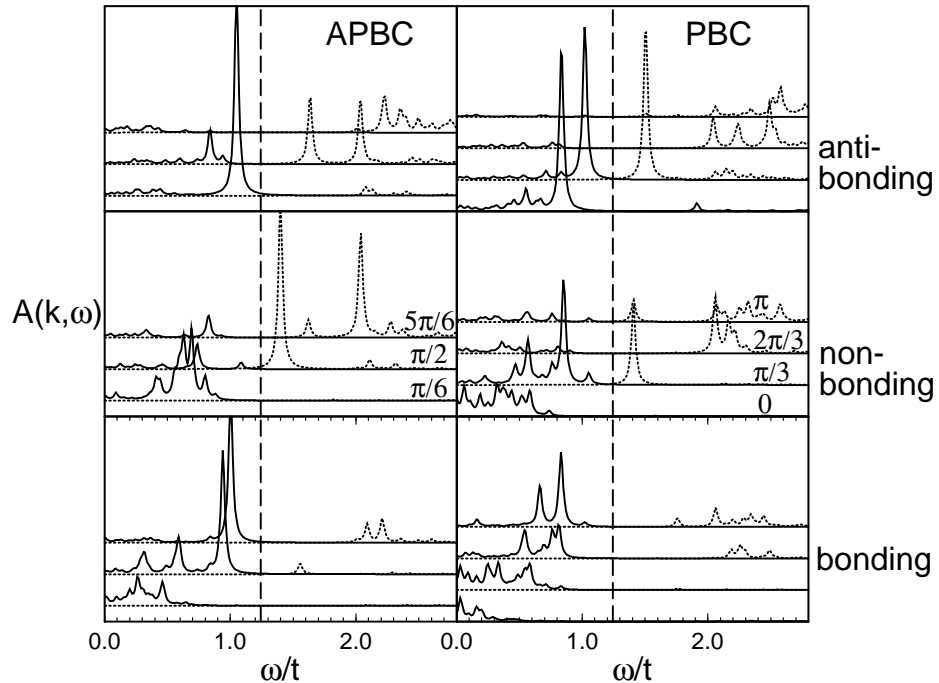}
}
\vspace{2mm}
\caption{Spectral functions for a (3$\times$6)-site
isotropic $t-J$ ladder with two holes and
$J/t$=0.5. The poles have been given a finite width of $\eta = 0.02t$,
and the chemical potential is indicated by the dashed line. 
APBC (PBC) were used in the left (right) panel.}
\end{figure}

Clearly all three bands ($b$, $ab$, and $nb$) are partially occupied.
This is illustrated by calculating the momentum distribution function,
$\rm n_{\lambda}(k) = \int^{\mu} d\omega A_{\lambda}(k,\omega )$.
The results for $\rm n_{\lambda}(k)$ are shown in Fig. 11. These show
drops as k increases which one can interpret as giving estimates for
$\rm k_{F,\lambda}$, but in a small system one cannot draw conclusions
about the actual behavior for $\rm k \sim k_{F,\lambda} $. The case of PBC 
has a different shell structure. The LL + ISL state is now a state with 
total momentum $\pm 2\pi/3$, with the 4 electrons in the $nb$-channel
occupying k=0 (2 electrons) and k=$\pm \pi/3$ (2 electrons)\cite{footnote3}.
The values of $\rm n_{\lambda}(k)$ for PBC in the LL + ISL state are also
included in Fig. 11, and also confirm the interpretation of a partial
filling of all three bands.

\begin{figure}[htb]
\vspace{2mm}
\centerline{\epsfxsize 11cm
\epsffile{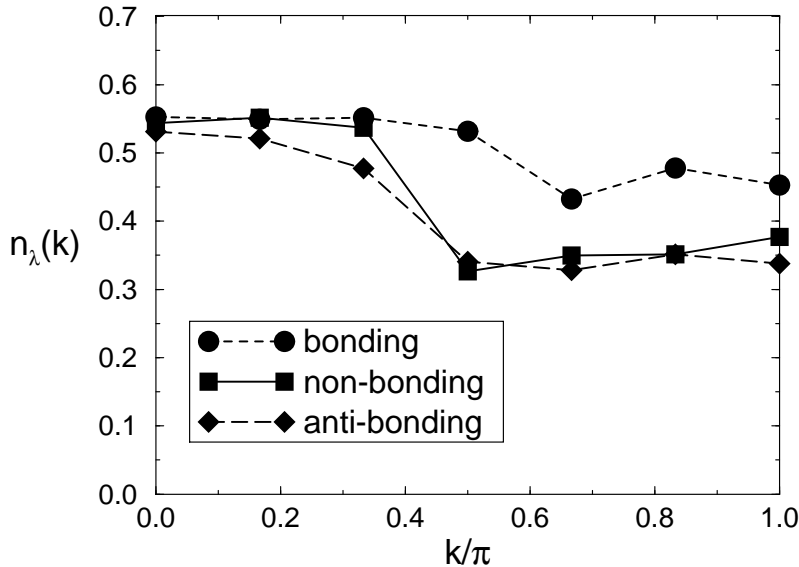}
}
\vspace{2mm}
\caption{Momentum distribution function for a (3$\times$6)-site
isotropic $t-J$ ladder with two holes and
$J/t$=0.5. Results with APBC and PBC were combined.}
\end{figure}

However, at a true Fermi surface in a bulk system one must be able
{\it to add} as well as to remove electrons at the chemical potential
(or Fermi energy). The strong correlation condition clearly influences this
part of the spectrum much more strongly, since as we pass to the undoped
Mott insulator electron addition is totally forbidden (or allowed only on
paying the Mott correlation energy gap), whereas in photoemission electrons
can always be removed. With this in mind we examine the electron 
addition part of the spectrum which corresponds to a transition to a 
one-hole state. From Fig. 10 we see that the main weight at low energy is 
in the $nb$($\pm \frac{\pi}{2}$)-channel. Clearly this corresponds to a
transition from the two-hole LL + ISL-groundstate to the one-hole
LL + ISL-groundstate, and confirms that we have a Fermi surface in the
odd-parity LL-channel. Next we look at the even-parity channels which form an 
ISL in the groundstate. Therefore we should expect that then electron 
addition in these channels at low energy will be forbidden. In fact, if we
look at the $b$-channel, then this behavior is clear, and there is no 
Fermi surface in this channel. On the other hand, in the other
even-parity channel, $\lambda = ab$, we find two relatively strong peaks
at $\rm k=\pm \frac{\pi}{2}$. These values actually correspond to a 
transition not to the minimum energy state of one hole in this channel
but to an excited state. We interpret these peaks as sidebands, involving the
creation of a S=1 odd-parity magnon with $k=0$ or $\pi$ in addition.
Clearly it is too strong a statement to say that all processes which add
an electron to a LL + ISL-state with even parity are forbidden, since
sidebands in which an electron is added in the odd-parity LL channel and
simultaneously odd-parity S=1 magnons are created can have a total parity
which is even and a total spin of S=1/2. We interpret the
weight in the $ab$-channels then in terms of such processes which in
turn may be enhanced by the finite size effects associated with these small
clusters. The behavior of the PBC case is analogous. Here the main weight
at low energy is the $nb$ (k=$\pi/3$) channel, corresponding also to a
transition to the one-hole LL + ISL state. However, again there is weight
in the antibonding channel which we interpret also as a higher energy
sideband and a transition to a one-hole LE state. This is strongly admixed
due to the enhanced stability of this state which has a filled shell 
character in the $nb$-channel as we discussed earlier.

In conclusion, the single particle spectra in the LL + ISL state show a
strong asymmetry between removing and adding electrons, and it is the
latter process, which unfortunately is not easy to detect experimentally,
where the effects of the Fermi surface truncation on approaching the Mott
insulating state are most evident.

\section{Analysis by Mean Field Theory}

In this section we analyze the properties of the three-leg ladder
using the
mean field description in the same spirit as previously 
done for the two-leg ladder \cite{SRZ}. For this purpose we introduce
spinon and holon 
operators, $f$ and $ b $ respectively, by replacing the electron
creation and annihilation operators in the following way \cite{KOTLIAR},

\begin{equation}
c^{\dag}_{i,\nu, \sigma} = f^{\dag}_{i, \nu, \sigma } b_{i,\nu} \qquad {\rm
  and} \qquad  c_{i,\nu,\sigma } =  b^{\dag}_{i,\nu} f_{i,\nu, \sigma }
\end{equation}
which lead to the local constraint $ \sum_{\sigma}
f^{\dag}_{i,\nu,\sigma} f_{i,\nu,\sigma} +
b^{\dag}_{i,\nu} b_{i,\nu} =1 $. The Hamiltonian can be reformulated in these
operators, and can then be decoupled by introducing mean fields. The
constraint is included by adding a term with Lagrange multipliers.
Since this treatment is quite standard we do not go into details here.
The system is not translationally
invariant along the rung, so that the mean fields depend on the position of the
bonds, resulting in 10 independent mean fields: 6 hopping and
3 pairing mean fields, defined on nearest neighbor bonds $ (j,\nu;j',\nu') $,

\begin{equation} \begin{array}{l}
\chi_{j,\nu;j',\nu'} = \frac{1}{2} \sum_{\sigma}  \langle f^{\dag}_{j,\nu,
  \sigma} f_{j',\nu', \sigma } \rangle \\ \\
B_{j,\nu;j',\nu'} = \langle b_{j,\nu} b^{\dag}_{j',\nu'} \rangle \\ \\
\Delta_{j,\nu;j',\nu'} = \langle f_{j,\nu \downarrow} f_{j', \nu'
  \uparrow} \rangle \\
\end{array}
\end{equation}
and a uniform Lagrange
multiplier $ \mu $.
We denote the mean fields along the outer two legs and the middle leg
with the index 1 and 2, respectively, while the rung mean fields have
the index 3. In order to avoid ambiguities we have to define the
direction of the bond mean fields. We give here the convention using
the example of $ \chi $,

\begin{equation} \begin{array}{l}
\chi_1 = \frac{1}{2} \sum_{\sigma} \langle f^{\dag}_{j,\nu, \sigma}
f_{j+1,\nu,\sigma} \rangle \qquad  \nu = 1,3 \\ \\
\chi_2 = \frac{1}{2} \sum_{\sigma} \langle f^{\dag}_{j,2, \sigma}
f_{j+1,2,\sigma} \rangle \\ \\
\chi_3 = \frac{1}{2} \sum_{\sigma} \langle f^{\dag}_{j,2, \sigma}
f_{j,\nu,\sigma} \rangle \qquad   \nu = 1,3 \\
\end{array} \end{equation}
The same convention is applied to the mean fields $ B $ and $ \Delta $.
The parity with respect to the reflection operator $R$
allows us to separate the mean field
Hamiltonian into the even (+)  and odd ($-$) parity channels,

\begin{equation} \begin{array}{ll}
H_{{\rm mf}}= & H^{(b)}_+ + H^{(b)}_- + H^{(f)}_+ + H^{(f)}_- \\ & \\
& + L [\mu +\frac{J}{2}
(2 \chi^2_1 + \chi^2_2 + 2 \Delta^2_1 + \Delta^2_2) + J' ( \chi^2_3
+ \Delta^2_3) \\ & \\ & + 2 t ( B_1 \chi_1 + B_2 \chi_2 ) + 4 t' B_3 \chi_3 ]
\end{array} \end{equation}
We define the corresponding combinations for the operators,
operator combinations,

\begin{equation} \begin{array}{ll}
f_{k \pm ,\sigma} & =  \sqrt{\frac{1}{2L}} \sum_j
(f_{j,1,\sigma} \pm  f_{j,3,\sigma}) e^{ikr_j} \\
b_{k \pm} & = \sqrt{\frac{1}{2L}} \sum_j (b_{j,1} \pm
b_{j,3}) e^{ikr_j} \\
\end{array} \end{equation}
with the momentum $ k $ along the legs of the ladder. Then
we can write the four terms of the Hamiltonian (in Nambu space for the
spinons) as

\begin{equation} \begin{array}{ll}
H^{(b)}_- & = (-4 t \chi_1 \cos k - \mu) b^{\dag}_{k-} b_{k-} \\ \\
H^{(b)}_+ & = (b^{\dag}_{k+}, b^{\dag}_{k,2}) \left[ \begin{array}{cc}
-4 t \chi_1 \cos k - \mu & - 2 \sqrt{2} t' \chi_3 \\
-2 \sqrt{2} t' \chi_3 & -4 t \chi_2 \cos k - \mu \\ \end{array}
\right] \left( \begin{array}{c} b_{k+} \\ b_{k,2} \end{array} \right)
\\ \\
H^{(f)}_- & = \sum_{\sigma} (f^{\dag}_{k-, \sigma}, f_{-k-,-\sigma}) \left[
  \begin{array}{cc} -(2 t B_1 + \frac{3}{2} J \chi_1)\cos k - \mu & - \frac{3}{2}
    J \Delta_1 \cos k \\   - \frac{3}{2} J \Delta_1 \cos k &
    (2 t B_1 + \frac{3}{2} J \chi_1) \cos k + \mu \end{array} \right] \left(
  \begin{array}{c} f_{k-,\sigma} \\ f^{\dag}_{-k-,-\sigma}
  \end{array} \right)
\\ \\
H^{(f)}_+ & = (f^{\dag}_{k+ \uparrow}, f^{\dag}_{k,2, \uparrow}, f_{-k+
  \downarrow}, f_{-k,2, \downarrow})  \left[
\begin{array}{cc} \hat{\xi}_k & \hat{\Delta}_k \\ \hat{\Delta}_k & -
  \hat{\xi}_{-k} \end{array} \right] \left( \begin{array}{c}
f_{k+ \uparrow} \\ f_{k,2, \uparrow} \\ f^{\dag}_{-k+
  \downarrow} \\ f^{\dag}_{-k,2, \downarrow} \end{array} \right)
\\
\end{array} \end{equation}
where $ \hat{\xi}_k $ and $ \hat{\Delta}_k $ are $ 2 \times 2
$-matrices of the form

\begin{equation} \begin{array}{l}
\hat{\xi}_k = \left[ \begin{array}{cc} -(2t B_1 + \frac{3}{2} J
    \chi_1) \cos k -
    \mu & - 2 \sqrt{2}t' B_3 - \frac{3 \sqrt{2}}{4} J' \chi_3 \\ & \\
     - 2 \sqrt{2} t' B_3 - \frac{3 \sqrt{2}}{4} J'
        \chi_3 & - (2 t B_2 + \frac{3}{2} J \chi_2) \cos k - \mu \end{array}
    \right] \\ \\
\hat{\Delta}_k = \left[ \begin{array}{cc} - \frac{3}{2} J \Delta_1
    \cos k & - \frac{3 \sqrt{2}}{4} J' \Delta_3 \\ & \\ - \frac{3
      \sqrt{2}}{4} J' \Delta_3 & - \frac{3}{2} J \Delta_2 \cos k
  \end{array} \right] \\
\end{array} \end{equation}
Note that we have neglected the terms with $ - n_i n_j/4 $ in the
exchange term, because in the mean field calculation they tend to
artificially enhance the tendency towards pairing and favor a flux phase
close to half-filling.

We can now solve the single-particle problem of $ H_{{\rm mf}} $ and
determine the mean fields self-consistently. For
the groundstate we obtain the self-consistent equations by minimizing
the groundstate energy of $ H_{{\rm mf}} $ in Eq.(4) with respect to
all mean fields. In this description the holons are Bose-condensed so
that $ \sum_{\nu=1,2,3} \langle b^{\dag}_{j,\nu} b_{j,\nu} \rangle = 3
\delta $ where $ \delta $ is the doping concentration. In the
following, we will use the same parameters as used in the numerical
simulations, $ t = t' = 2 J = 2 J' $.

\subsection{Half filling}

For half-filling, $ \delta=0 $, there are no holons and the Hamiltonian
of the spinons is invariant under the following SU(2) transformation

\begin{equation}
\left( \begin{array}{c} f^{\dag}_{j,\nu,\sigma} \\ f_{j,\nu,-\sigma}
    \end{array} \right) \to \hat{U} \left( \begin{array}{c}
      f^{\dag}_{j,\nu,\sigma} \\ f_{j,\nu,-\sigma }
    \end{array} \right) = \left( \begin{array}{c}
      f'^{\dag}_{j,\nu,\sigma} \\ f'_{j,\nu,-\sigma }
    \end{array} \right)
\end{equation}
which reflects the constraint at half-filling. The absence of an 
up-spin corresponds to the presence of a down-spin and vice versa
\cite{affleck}. Note that due to the
SU(2) invariance of the Hamiltonian the self-consistent solution of
the mean field is not unique, but $ \hat{U} $ corresponds to a
rotation in the mean field space $ \{ \chi_{1,2,3} , \Delta_{1,2,3} \} 
$ which leaves the groundstate and the excitation spectrum
unchanged. 

\begin{figure}[htb]
\centerline{\epsfxsize 8 cm
\epsffile{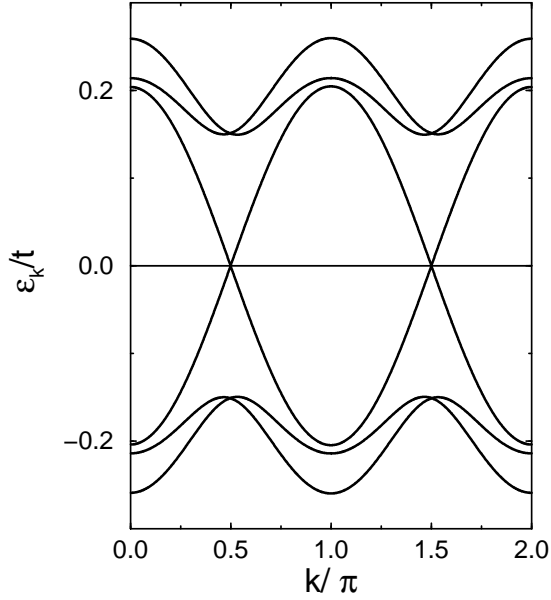}
}
\vspace{2mm}
\caption{Spinon bands at half filling obtained by the mean field
  theory. All states below the chemical potential (located at
$\rm \epsilon_k = 0$) are occupied
  by a spin-up and a spin-down spinon. The gapped bands correspond to the
  even-parity states, while the gapless bands in the center are in
  the odd-parity channel.}
\end{figure}

In this formulation various properties of the three-leg
ladder can be interpreted at least on a qualitative level.
The Hamiltonian yields six distinct
bands in Nambu space for the spinons (Fig. 12 ). For the symmetric
channel the spinons have a gapped spectrum where the lower two bands
are completely filled. On the other hand, the
antisymmetric channel has gapless excitations which give a spectrum
analogous to that of a single chain. It is now easy to compare
the spin excitation spectrum of the mean field and the exact
diagonalization. For the mean field case the excitation corresponds to
a particle-hole excitation of the spinon gas. The gapless spectrum is
that of a single Heisenberg chain and is exactly what we obtain for
the same type of mean field treatment of a single chain. The other
bands are gapped and describe a spin liquid state coexisting with
the gapless system. The agreement with the numerical simulation is
qualitatively very good.

\subsection{A single hole}

We now insert a single hole into the half-filled system, i.e. we
remove one spinon and add one holon. This immediately breaks the SU(2)
symmetry, because the groundstate energy of the holon has to be
minimized which yields a constraint on the hopping mean fields $ \chi_i
$. Three holon bands appear, two even parity bands and one odd parity 
band, with energies given by
\begin{equation} \begin{array}{ll}
\epsilon_{k+} = & -2t (\chi_1 + \chi_2) \cos k - \mu \\ & \\ & \pm
\sqrt{4 t^2
  (\chi_1 - \chi_2)^2 \cos^2 k + 8 t^2 \chi^2_3 } \\ \\
\epsilon_{k-} = & - 4 t \chi_1 \cos k - \mu .
\end{array} \end{equation}
Obviously the lowest holon state can be found in the lower of the two
even-parity bands. Thus, the single hole groundstate can be obtained by
removing a spinon at the Fermi level of the odd-parity band and by
inserting a holon at bottom of the lower even-parity holon band. This
results in a change of the parity of the groundstate compared with
that of the half-filled case, in agreement with the numerical
result.

\begin{figure}[htb]
\centerline{\epsfxsize 8cm
\epsffile{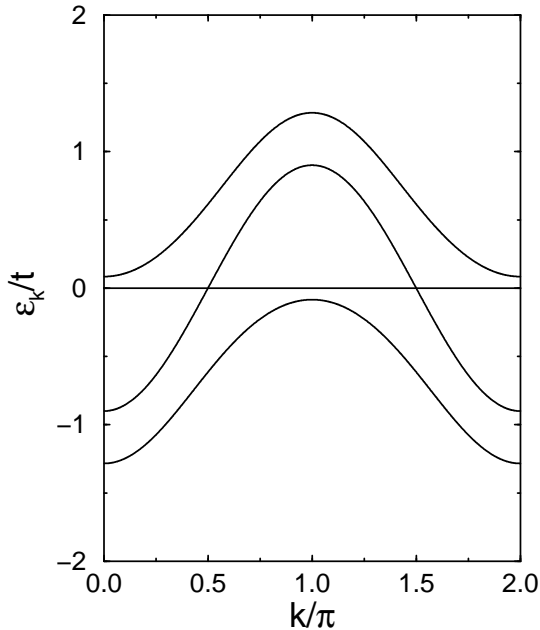}
}
\vspace{2mm}
\caption{Holon bands for a single hole obtained by the mean field
  theory. The outer bands are the even-parity bands (bonding and
  anti-bonding) and the center band has odd parity (non-bonding).}
\end{figure}

The single particle spectrum is incoherent,
because the excitations are composed of the annihilation of a spinon
and the creation of a holon. We define the single hole Green's function
as a matrix

\begin{equation}
G_{\nu \nu'} (k,\omega) = - \int d t e^{-i \omega t} \langle T (
c^{\dag}_{k,\nu, \sigma}(t) c_{k,\nu', \sigma}(0) \rangle
\end{equation}
where the indices $ \nu $ and $ \nu' $ denote the three legs of the
ladder ($T$ the time-ordering operator). Decomposing the $ c $-operators
into spinon and holon parts we obtain the
convolution

\begin{equation}
G_{\nu \nu'} (k, \omega )= \frac{1}{L} \sum_q \int d \omega' G^f_{\nu \nu'}
(k+q, \omega + \omega') G^b_{\nu \nu'} (q, \omega')
\end{equation}
with

\begin{equation} \begin{array}{l} \displaystyle
G^f_{\nu \nu'} (k, \omega) = - \int dt e^{-i \omega t} \langle T
( f^{\dag}_{k,\nu, \sigma}( t)
f_{k,\nu', \sigma } (0) ) \rangle  \\ \displaystyle
G^b_{\nu \nu'} (k, \omega)  = - \int dt e^{-i \omega t} \langle T ( b_{k,\nu }(t)
b^{\dag}_{k,\nu' } (0) ) \rangle
\end{array}
\end{equation}

\begin{figure}[htb]
\centerline{\epsfxsize 8cm
\epsffile{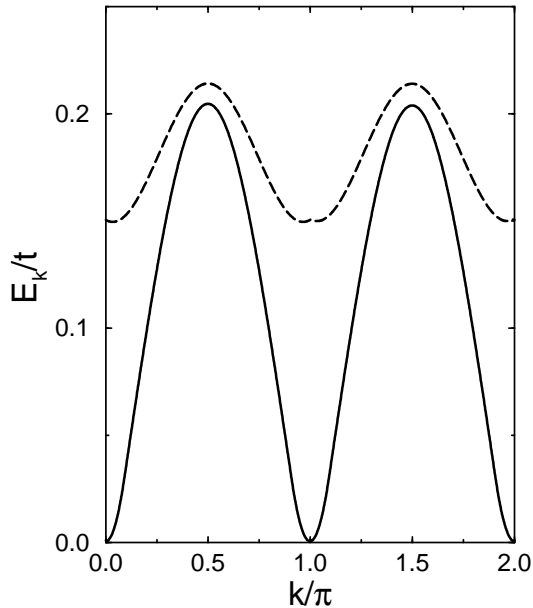} 
}
\vspace{2mm}
\caption{Low-energy spectrum for a single hole obtained by the mean field
  theory.}
\end{figure}
The spectrum is obtained by diagonalizing the Green's function with
respect to the leg indices. We show in Fig. 14 the lowest one-hole states
for given momenta and parity analogous to the numerical results in
Fig. 5. The spectrum changing the parity with respect to the
groundstate of the half-filled system has gapless excitations at
momentum $ k =0 $ and $\pi $. This corresponds to the LL
component of the three leg ladder. The states which keep the parity of
the groundstate,
however, are gapped with minima at $ k=0 $ and $ \pi $. In our mean field
calculation the latter excitations lie above the LL
spectrum, in contrast to the numerical result. The reason is that in
the mean field treatment these excitations are almost exclusively
carried by the spinon part, i.e. we remove a spinon in the lower
even-parity spinon bands, while the holon remains in the
lowest even-parity band. In the numerical calculation this spectrum lies clearly lower
and cannot be identified directly with spinon excitations which are
higher in energy. Therefore,
we conclude that our mean field calculation overestimates the splitting
of the holon bands. Nevertheless, we can interpret both types of excitations
consistently with the numerical calculations. In particular, we
emphasize that the gapless spinons are in both treatments in odd-parity
channel giving a consistent picture of the symmetries of the ground
state and the low-lying excitations.

\subsection{Finite doping}

The analysis of our numerical data led us to the conclusion that there would
be a finite range of doping close to half-filling where the LL
state would coexist with the ISL.
Beyond a critical
doping $ \delta_c $ the ISL would be doped and the whole
system would turn into a
LE where all channels open a spin gap. We might hope
to describe this behavior within
our mean field picture. However, in this treatment
the mean field LL state is immediately unstable against
the formation of a LE state, and there is no obvious
transition between two different phases.

As soon as we introduce a finite concentration of holes in the system,
the holons are described by a Bose-condensate in the mean field
groundstate. Consequently, the single-particle
Green's function consists of a coherent and an incoherent part due to
the presence of the condensate,

\begin{equation}
G_{\nu \nu'} (k, \omega) = C_{\nu \nu'} G^f_{\nu \nu'}(k, \omega) +
G^{{\rm inc}}_{\nu \nu'} (k, \omega)
\end{equation}
where $ C_{\nu \nu'} $ is a constant proportional to the doping
concentration $ \delta $. Thus, we find conventional Fermi liquid
quasiparticles in the single particle spectrum. This has implications
on the pairing in the doped ladder. The exchange term in
the $ t $-$J$ model allows to scatter spin-singlet
pairs of electrons between the even-parity and odd-parity channel. The relevant terms
are derived from the exchange terms along the two outer legs,
$ J \sum_1 ({\bf S}_{i,1} \cdot {\bf S}_{i+1,1} +  {\bf S}_{i,3} \cdot
{\bf S}_{i+1,3} ) $, and become in electron operator formulation,

\begin{equation} \begin{array}{rl}
H_{+-} =  \sum_{k,k',q} \sum_{\sigma = \uparrow, \downarrow}
v_{k,k',q} &
[c^{\dag}_{k+q +, \sigma} c^{\dag}_{k'+,
  -\sigma} c_{k -, -\sigma} c_{k'+q -, \sigma} \\ & + c^{\dag}_{k+q -, \sigma}
c^{\dag}_{k'-, -\sigma} c_{k +, -\sigma} c_{k'+q +, \sigma}].
\end{array} \end{equation}
with $ v_{k,k',q} = - J ( \cos q + \frac{1}{2} \cos (k-k')) $. The
even-parity channel corresponds to the ISL phase which is
not populated by holes at small doping.
However, via virtual scattering of hole pairs
into the even-parity channel an effective pairing
interaction for the holes in the
odd-parity channel is generated, in lowest order given by

\begin{equation} \begin{array}{ll}
V_{k,k'} & =  - 2 i \sum_{k''} \int d \omega  v_{k,k'',k+k''}
G^f_{++}(k'',\omega) G^f_{++} (-k'', - \omega) v_{k',k'',k'+k''} \\ &
\\ & \displaystyle
= - \frac{J^2}{2} i \sum_{k''} \int d \omega (9 \cos k \cos k'
\cos^2k'' + \sin k \sin k' \sin^2 k'') G^f_{++}(k'',\omega) G^f_{++}
(-k'', - \omega) \end{array}
\end{equation}
where $ G^f_{++} (k, \omega) $ is the spinon Green's function in the
even-parity channel corresponding to $ G^f_{++} = \frac{1}{2} (G^f_{11}
+ G^f_{13} + G^f_{31} + G^f_{33} ) $ from Eq.(15). In weak coupling
theory this leads to the instability equation (ladder approximation)
in the odd-parity channel,

\begin{equation}
1 = - k_B T \sum_{k,n} V_{k,k} G_{--} (k,\omega_n) G_{--}
(-k, - \omega_n)
\end{equation}
for finite temperature ($ G_{--} = G_{11} - G_{13} - G_{31} + G_{33}
$, and $ \omega_n $ are the fermionic Matsubara frequencies). Since the
Green's function has a
standard quasiparticle pole as shown above in Eq.(16), we find a usual
Cooper instability and a
non-vanishing transition temperature for any finite coupling
and density $ \delta $ (note that $ G $ is proportional to $ \delta
$). Thus, holes in the odd-parity channel would be paired in the
groundstate. This is in contrast to the expectation for a LL
where a critical coupling strength exists below which the
LL remains stable \cite{BALATSKY}. Thus, this
instability of the LL for any non-zero
concentration of holes is a deficiency of the mean field description.
On the other hand, it clearly indicates the trend towards pairing due
to the coupling of the LL to the insulating or doped spin liquid which
occupies a part of the spectrum.
The RVB correlation of the ISL provides a pool of
``preformed pairs'' which triggers the instability of the LL
towards the LE \cite{GIL}.

\begin{figure}[htb]
\centerline{\epsfxsize 8cm
\epsffile{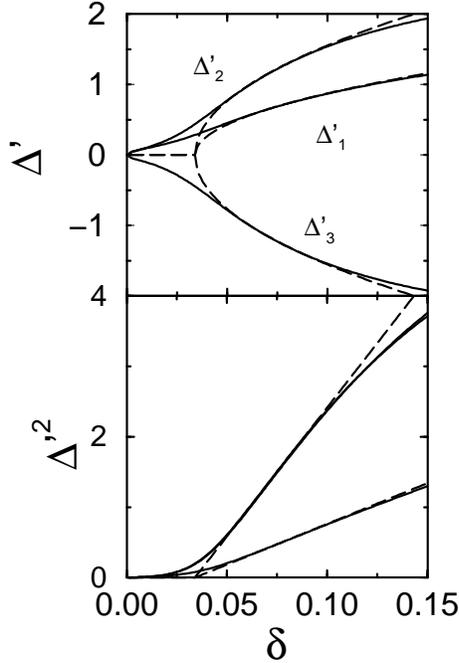}       
}
\vspace{2mm}
\caption{The superconducting mean fields for finite doping. The solid
  lines correspond to the mean field calculation. The dashed lines
  are fits of the crossover assuming an underlying $ \sqrt{\delta -
    \delta_c} $-dependence of $ \Delta' $ with $ \delta_c \approx
  0.034 $. In the lower figure a linear fit is possible for $
  \Delta'^2 $.}
\end{figure}

Although there is no obvious (non-zero) critical concentration where the
character of the groundstate changes within the mean field treatment,
we observe an interesting crossover between two regimes in the BCS
mean field for the holes,

\begin{equation}
\Delta'_{ij} = \langle c_{i \uparrow} c_{j \downarrow} \rangle
= \langle b^{\dag}_i b^{\dag}_j \rangle \langle f_{i \uparrow} f_{j
  \downarrow} \rangle = B_{ij} \Delta_{ij} .
\end{equation}
The behavior of $ \Delta' $ and $ \Delta'^2 $ as a
function of $ \delta $ is shown in Fig. 15. The crossover indicates
a critical concentration $ \delta_c \approx 0.034 $ (for $ t=2J $)
found by extrapolation of the square root dependence of $
\Delta' $ ($ \propto \sqrt{\delta - \delta_c} $). For $ 0 <
\delta \ll \delta_c $ $ \Delta' $ is rather small as can be
understood from Eq.(15) and (16). For small doping the dominant
contribution to $
V_{k,k'} $ originates from $ k'' \approx \pm \pi/2 $ the momenta for
the spin gap in the even-parity channel and the Fermi points in for the odd-parity
holes are close to $ \pm \pi/2 $ as well.  Therefore
the cosine-term in $ V_{k,k'} $ contributes only little.
This changes with increased doping where both the Fermi points and the
lowest even-parity spinon energies are located at momenta gradually shifting
away from $ \pm \pi/2 $. This leads to an effective enhancement of the
pairing interaction as an effect of doping which yields the pronounced crossover whose
symmetry aspects we discuss below. 
We expect that the same tendency is at work in the real system where
in the very small doping regime the effective interaction is
sufficiently reduced to avoid pairing in the LL. Only with
increased doping both the enhanced attractive interaction and the
density of states at the Fermi level drive the system into the paired
state.

Finally we would like to characterize this crossover by considering
the symmetry of the gap function in
momentum space. Thus, we introduce transverse momenta $ k_{\perp}
$ along the rung. The bonding, antibonding and non-bonding states as given
in Eq.(2) and (3) can be related to momenta $ k_{\perp} $ by using the
following form for the dispersion,

\begin{equation}
\epsilon_{k,k_{\perp}} = -2t(\cos k + \cos
k_{\perp} ),
\end{equation}
resulting from fixed boundary conditions along the rung. 
With Eq.(2) and (3) we find the correspondence: $ b \to
k_{\perp} = \pm \pi/4 $, $ nb \to k_{\perp} = \pm \pi/2 $ and $ ab \to
\pm 3 \pi/4 $. Then we obtain for the momentum dependent gap function,

\begin{equation} \begin{array}{ll}
\Delta'_{k, k_{\perp}} & = \langle c_{k,k_{\perp} \downarrow}
c_{-k,-k_{\perp} \uparrow} \rangle  \\ & \\
& = -(2 \Delta'_1 \sin^2 (k_{\perp}) + \Delta'_2 \sin^2(2 k_{\perp}) )
\cos k - \Delta'_3 \sin(2 k_{\perp})(\sin (k_{\perp}) + \sin (3
k_{\perp})).
\end{array} \end{equation}
We can use this form now to show the phase of the gap function in the
first Brillouin zone (BZ). We observe that the gap function is basically
positive along the $ k $-direction ($ | k_{\perp}| \ll \pi $)  and
negative along the transverse 
momentum direction $ k_{\perp} $ ($ |k| \ll \pi $). This is
essentially the structure of 
a ``$d_{x^2-y^2} $-wave'' pairing function. This is also reflected in
real space where the gap function along the legs is positive ($
\Delta'_1 , \Delta'_2 $) and along the rungs is negative ($ \Delta'_3
$) as shown in Fig. 15. It is interesting to compare the position
of the nodes in the two regimes separated by $ \delta_c $. For $ 0 <
\delta \ll \delta_c $ we find that the gap functions have practically
the same magnitude on all bonds. With this property Eq.(22) leads to
nodes which correspond exactly to the [110]-direction in the BZ and we might
consider this state as ``purely'' $ d_{x^2-y^2} $-wave like (see Fig. 16). On the
other hand, for the regime $ \delta > \delta_c $ the gap functions
have the relation $ \Delta'_1 < \Delta'_2 = - \Delta'_3 $ which
yields nodes clearly shifted away from the [110]-direction. We may
consider this as an admixture of an (extended) s-wave-like component to the
d-wave gap, although, strictly speaking, the underlying symmetries are
not present in the ladder system to justify the distinction between s- 
and d-wave pairing.

\begin{figure}[htb]
\centerline{\epsfxsize 8cm
\epsffile{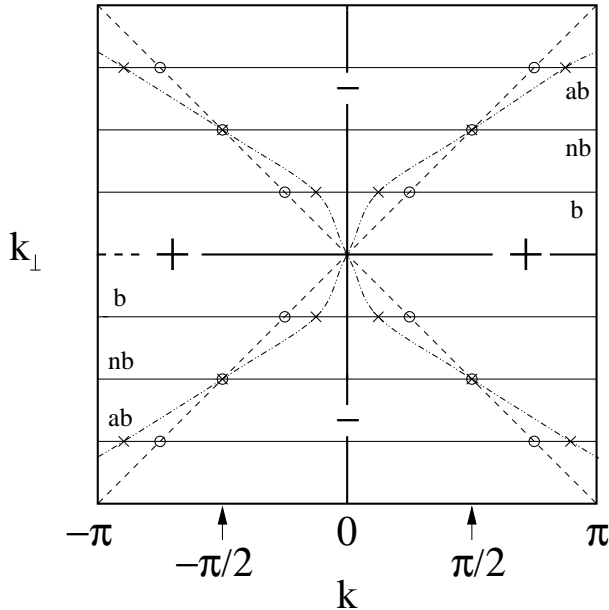}
}
\vspace{2mm}
\caption{Nodelines for the BCS gap, Eq. (22),  in the first Brillouin zone. 
The dashed line connecting the empty circles shows the nodelines for $ 
\delta = 0.01 $. The dotted-dashed line connecting the x-marks gives
the nodelines for $ \delta=0.06 $. We
define the transverse momenta $ k_{\perp} $ by $ b \to \pm \pi/4 $, $
nb \to \pm \pi/2 $, and $ ab \to \pm 3 \pi/4 $.  }
\end{figure}

The difference of the two regimes may therefore be interpreted in
the following way. The mean field  superconducting state in the small
doping regime is mainly carried by the even-parity channel and the odd-parity channel
participates weak through proximity. Contrary to the exact numerical
diagonalization, in the mean field treatment we
cannot avoid the population of the ISL by holes strictly even at very
small doping concentrations. The three-leg structure is
apparently wide enough to create a LE state with a 
gap structure which is approximatively
d-wave like. In the larger doping regime,
however, the LL of the odd-parity channel acquires its own hole pairing so
that an additional symmetry lowering occurs reflecting the 1D nature
of the LL and
generates an s-wave like additional component as seen in
the shift of the nodelines (Fig. 16). Since there is
no real symmetry distinguishing s- and d-wave from each other this
transition appears only as a crossover when the interaction among
holes become strong enough in the (odd-parity) LL. This behavior is again in
qualitative agreement with the interpretation obtained from our
numerical calculation.

\section{Conclusions}

The three-leg ladder is specially interesting because in a certain sense
it combines two quite different elements - an odd-parity channel which 
behaves like a single chain, and two even-parity channels which behave similar
to the two-leg ladder. This result is already foreshadowed in the undoped
limit where the low-energy degrees of freedom can be mapped onto a 
single Heisenberg S=1/2 chain with longer range interactions, and the
remaining transverse spin degrees of freedom have a substantial energy gap.
In this work, we show that the initially doped holes enter only this
odd-parity channel and form a Luttinger liquid, so that the system has 
two distinct components - the conducting Luttinger liquid in the odd-parity
channel, and the insulating spin liquid in the even-parity channel. This 
phase we denoted as LL + ISL, and the numerical evidence from the Lanczos
diagonalization of small clusters reported here and the DMRG calculations
of White and Scalapino\cite{white}, 
that such a phase exists up to a critical hole 
density, $\rm \delta_c$, is we believe quite clear. In addition, we introduced 
a mean field approximation scheme which gave similar but not identical results.
Initially, holes enter only the odd-parity channel which is gapless in the 
undoped system. However in the mean field approximation, a small gap develops
in the odd-parity channel upon doping. This - as we discussed - reflects
the inadequacy of the mean field description of the Luttinger liquid.

The LL + ISL phase has unusual properties. First of all, we note that the
different parity channels of the original Fermi surface behave quite 
differently, so that only in the odd-parity channel is there a Fermi surface.
The truncation of the Fermi surface in the partially occupied 
even-parity channels is not a 
consequence of a breaking of translational symmetry since the spin order
here is purely short range. Rather it is a consequence of the proximity to
the Mott insulating phase which in this channel is ISL. This 
truncation of some partially occupied bands
is a clear violation of Luttinger's theorem.
Usually if one approaches a Mott insulator which has AF order, then one may 
proceed through incommensurately ordered phases which progressively
truncate the Fermi surface, but do not violate Luttinger's theorem. 
However this option is not available if one approaches an RVB Mott 
insulator which is an ISL. However the example of the three-leg ladder shows
us that here also a partial truncation of the Fermi surface is possible,
but now it violates the Luttinger theorem.

This LL + ISL phase has certain features in common with a recent proposal
by Geshkenbein, Ioffe, and Larkin (GIL) for the underdoped spin gap normal
phase of the cuprates. They argue that the spin pairing implied by the 
spin gap did not cause immediately hole pairing. Instead they broke up 
the Fermi surface into two distinct parts - a fermionic part and a paired
bosonic part. The latter they argued should have infinite mass to prevent
conductivity from these bosons, and they associated this with the
van-Hove singularity. In our specific example, the spin pairs are also
insulating, but the origin lies in the proximity to the Mott insulating
phase. In both models, the coexistence of fermionic and bosonic degrees of
freedom leads to processes whereby a fermionic Cooper pair can scatter in and
out of the bosonic channels, which however lie at higher energy. In a LL,
there is no Cooper instability for an infinitesimal attraction, and a 
finite attraction is required for pairing. For this reason we believe
a LL + ISL phase is possible in the three-leg ladder.
By contrast, as we discussed above, the mean field description of the
odd-parity channel has a true Cooper instability due to the holons being
Bose condensed, and as a result hole pairing occurs at arbitrarily small
hole densities. In the case of two dimensions, where the Fermi surface 
channels or patches near to the saddle points, $(\pm \pi,0)$ and 
$(0,\pm \pi)$, first become paired and insulating, it is clearly crucial 
whether the remaining fermionic part of the Fermi surface has a Cooper
instability (as assumed by GIL) or not. A Cooper pairing instability leads
to hole pairing in the groundstate as in our mean field description.

The experimental examination of the Fermi surface evolution has been made
by ARPES (Angle-Resolved-Photoemission-Spectroscopy). However, as we have seen,
the inverse process would be more illuminating, but it is much more 
difficult to realize experimentally. In the electron addition process, as we
have shown, the Fermi surface truncation is very evident, although sidebands
do allow an electron addition on the even-parity channel which is in the
ISL state. Nonetheless the total weight at low energies will vanish as the
hole doping vanishes, and will be concentrated mainly in the fermionic
part of the Fermi surface. 

In the three-leg ladder at dopings beyond $\rm \delta_c$, the ISL is converted 
into a doped Luther-Emery liquid, and a LL + LE phase occurs. This phase 
will show hole pairing and power law correlations in the CDW and singlet
superconductivity channels. As the mean field approximation shows, the 
pairing is in an essentially d-wave channel. In the mean field theory, the
quantum phase transition at $\rm \delta_c$ appears as a crossover where the
hole pairing increases rapidly as the hole density increases.

We wish to thank V.B. Geshkenbein, R. Hlubina, L.B. Ioffe, P.A. Lee,
S.R. White, and D. Scalapino
for useful discussions,
and acknowledge
the Swiss National Science Foundation for financial support. In
particular, M.S. is grateful for the support by the Swiss National
Science Foundation through a PROFIL-fellowship.

%
%


\begin{references}
\bibitem{footnote2} As a consequence of the Marshall theorem, Heisenberg
chains and odd-leg ladders with L= 2, 6, 10, ... have a groundstate
wavevector k=$\pi$, while systems with L= 4, 8, 12, ... have k=0.
\bibitem{footnote3} Note that this 2-hole LL + ISL state for PBC lies
slightly higher in energy than the 2-hole LE state which has total momentum
k=0, and a filled shell configuration in the $nb$-channel, i.e. 6 electrons
with k=0 and k=$\pm \pi/3$ states fully occupied. We ascribe the extra
stability of the LE state to this favorable filled shell structure.
\end{references}
\end{document}